# Methyne Capping in the Boron Buckyball : A Viable Possibility

Jules Tshishimbi Muya,[a] Minh Tho Nguyen[a] and Arnout Ceulemans*[a]

*Department of Chemistry, and Institute for Nanoscale Physics and Chemistry (INPAC), University of Leuven, Celestijnenelaan 200F, B-3001 Leuven, Belgium*

**We report the electronic structure of methyne boron buckyballs $B_{68}(CH)_{12}$ and $B_{72}(CH)_8$ obtained by substituting respectively 12 and 8 boron cap atoms by methyne CH groups on the boron buckyball $B_{80}$. DFT calculations and minimization techniques have been employed to characterize the structural and electronic properties of endo and exo isomers of these molecules in $T_h$ symmetry. A vibrational frequencies analysis predicts that only endo-$B_{72}(CH)_8$ corresponds to a true minimum on the potential energy hypersurface with a cohesive energy similar to the boron buckyball. The viable existence of this carboron buckyball opens new perspectives for a synthesis of large boron clusters.**

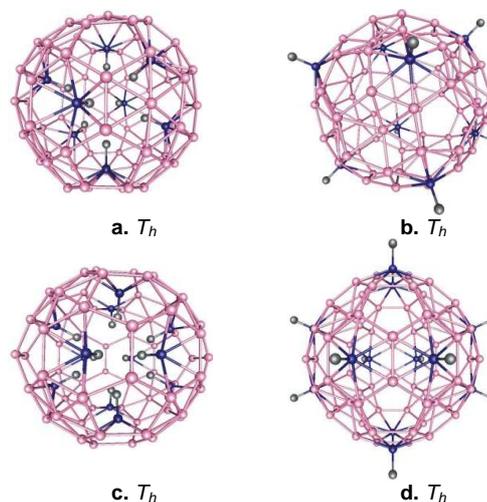

**Fig. 1** Optimized geometry of $B_{80-x}(CH)_x$ isomers calculated with DFT at B3LYP/SV(P) level: (a) endo-$B_{72}(CH)_8$, (b) exo-$B_{68}(CH)_{12}$, (c) endo-$B_{68}(CH)_{12}$ and (d) exo-$B_{72}(CH)_8$

$B_{80}$, the boron buckyball,[1] is probably the most interesting new molecule coming out of current quantum chemical studies. Actual synthesis still seems a remote possibility, but in the mean time relentless theoretical activity is going on predicting many new properties of this boron cluster. Theoretical studies revealed that the boron buckyball has a geometry which is slightly distorted from $I_h$ to $T_h$ symmetry[2,3] and the analysis of chemical bonding demonstrated a perfect match between the symmetries of the bonding orbitals of $B_{80}$ and the original buckminsterfullerene $C_{60}$.[4] Similar to $C_{60}$, $B_{80}$ can also condense to form a simple cubic (sc), face centred cubic (fcc) or body centred cubic (bcc) solid cluster. The fcc solid cluster appears more stable than the bcc lattice or the sc.[5,6] The electronic transport transmission in $B_{80}$ carried out by *ab initio* calculations is shown to be higher than in $C_{60}$ in the Fermi region.[7] A DFT study on alkali metal doped $B_{80}$ revealed that the $Na_{12}B_{80}$ and $K_{12}B_{80}$ molecules have high capacity to store hydrogen molecules up to 72 molecules and they are excellent candidates for hydrogen storage.[8]

It is well known that for every fullerene isomer $C_n$, there is a leapfrog fullerene $C_{3n}$ that has the same idealized point group and a closed-shell electronic structure.[9-10] In the same way a boron leapfrog transformation from $C_n$ to $B_{3n+n}$ clusters can be defined. In a recent communication, Yan *et al*. found that the leapfrog operation could be considered as a generic constructing scheme that can produce a large family of novel stable boron nanostructures.[11] The bonding analysis in $B_{80}$ shows that the capping borons centers donate their valence electrons to the π-bonding in the truncated icosahedral frame. This suggests the viability of an isoelectronic substitution of the caps by fragments containing 3 valence electrons. A likely candidate is the methyne fragment CH which also substitutes for boron in a multitude of carborane cages. Hexacoordinate carbon has also been suggested by Exner and Schleyer as electron source for an aromatic boron rings.[12]

Initial studies on a fully substituted boron buckyball where all 20 cap atoms were replaced by methyne did not yield stable structures, so we have focussed our study on partially substituted isomers with either 8 or 12 caps replaced by CH groups. These isomers can be realised in $T_h$ symmetry, which also corresponds to the lowest energy structure of $B_{80}$ itself.

The methyne groups are oriented in the direction of the radius vector from the centre of the cluster, with the hydrogen either inside or outside of the cage. We will refer to this as endo and exo orientations respectively. Hence four $B_{80-x}(CH)_x$ structures (x=8,12) were examined, as indicated in Fig.1.

These methyne boron fullerenes are not isoelectronic to $B_{80}$ but have the same frame and open a way to extend the boron buckyball cage to large derivatives boron fullerenes.

A full geometry optimization was performed using Gaussian 03 Revision D02[13] and TURBOMOLE V-5-8-0 program packages.[14] The hybrid functional Becke's three-parameter (B3) incorporating the exact exchange functional in combination with Lee, Yang, and Parr's (LYP) correlation functional is used with the split-valence plus polarization SVP basis set. To examine that the optimized structures have reached the global minimum energy at the potential energy surface, we have analysed the vibration frequencies of these isomers.

The $T_h$ optimized structures of those four isomers, endo-$B_{72}(CH)_8$, endo-$B_{68}(CH)_{12}$, exo-$B_{72}(CH)_8$ and exo-$B_{68}(CH)_{12}$ calculated at B3LYP/SVP level are shown in Fig.1.

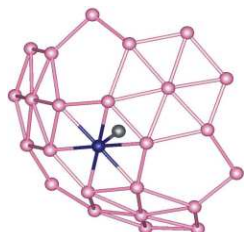

**Fig.2** Isolated methyne group in $B_{72}(CH)_8$
The carbon atom at the middle of the scheme in blue color is surrounded by pentagons and hexagons and linked to a H atom

The two $B_{72}(CH)_8$, isomers have all 8 carbons isolated by 3 pentagons and 3 hexagons (Fig.2) whereas in $B_{68}(CH)_{12}$ six pairs of adjacent hexagons carry methyne caps.

Of the four structures only the endo-$B_{72}(CH)_8$ was found to be stable in $T_h$ symmetry. Its softest vibrational mode occurs at 150 cm$^{-1}$ at the B3LYP/SVP level, and corresponds to the typical quadrupolar squashing mode of a vibrating sphere.[15] The three other isomers are metastable. The orientation of methyne groups, their number and localization on the buckyball cage seem to play an important role in the thermodynamic stability of the $B_{80-x}(CH)_x$ compounds. The calculated HOMO-LUMO gap, the cohesive energy and the number of imaginary frequencies are listed in Table 1.

**Table 1.** Calculated Total Energies and cohesive energies of $B_{80-x}(CH)_x$ with x=8 ,12 in $T_h$ symmetry at B3LYP/SVP level

| Isomers | Energy (in a.u) | HOMO-LUMO/(eV) | Cohesive energy/(eV) | N |
|---|---|---|---|---|
| Endo-72 | -2094.42 | 2.60 | -3.93 | 0 |
| Exo-72 | *-2094.32* | 1.82 | *-3.89* | 11 |
| Endo-68 | *-2149.66* | 2.07 | *-3.41* | 12 |
| Exo-68 | *-2149.08* | 2.08 | *-3.24* | 29 |
| $B_{80}$ | -1984.67 | 1.94 | -4.19 | 0 |

The values in italic of the total energy are not scaled with ZPE.
N represents the number of imaginary frequencies calculated at STO-3G level.

The only stable isomer endo-*72* has the highest HOMO-LUMO gap and the cohesive energy obtained at B3LYP/SVP is close to the cohesive energy of $B_{80}$ at the same level.

**Table 2.** Comparison of bond distances of $B_{80-x}(CH)_x$ isomers and others compounds such $B_{80}$ and carboranes molecules

| Molecules | $R_{B-B}$ (Å) | $R_{B-C}$ (Å) | $R_{C-H}$ (Å) | BCH angle |
|---|---|---|---|---|
| Endo-72 | 1.63-1.84 | 1.82-1.84 | 1.10 | 115.1-115.8 |
| $B_{80}$[1] | 1.65-1.76 | - | - | - |
| *1,12-$C_2I_2B_{10}H_{10}$*[16] | 1.77-1.78 | 1.72-1.74 | 1.10 | 115.2-118.5 |
| $C_8H_{16}B_{10}$[17] | 1.76-1.80 | 1.68-1.74 | | 106.3-108.3 |

We have done an additional single point calculation with a large basis set 6-31G(d). The energy of endo-72 calculated at B3LYP/6-31G(d) level is –2097.1652664 Hartrees and yields a cohesive energy of –4.86 eV approaching the cohesive energy of $B_{80}$ at the same level –5.17 eV reported in ref 2-7.

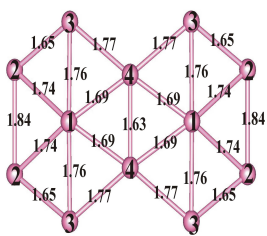

**Fig.3** Types of boron atoms and B-B bond lengths (Å)

Bond lengths are in the same range as the bond lengths found in compounds such as boron buckyball $B_{80}$, carboranes and different viable planar hexacoordinate carbon molecules suggested by Ito *et al.*[18] (Table 2).

In the optimized geometry of endo-72 molecule there are four different types of boron atoms present (see Fig.3).

Typically the two adjacent cap atoms of type B1 approach the common edge in between them, so as to form together with the B4 atoms a four-centre bond. The B-C bond lengths are more uniform. The carbon itself shows a large pyramidal distorsion towards the inside of the cage, with a BCH angle of about 115 degrees.

**Table 3.** Charge distribution per atom for differents atoms

| Atoms type | Charge per atom | Atoms number | Total cahrge |
|---|---|---|---|
| C | -0.54 | 8 | -4.34 |
| H | 0.25 | 8 | 2.00 |
| $B_1$ | 0.15 | 12 | 1.764 |
| $B_2$ | -0.002 | 24 | -0.04 |
| $B_3$ | 0.06 | 24 | 1.37 |
| $B_4$ | -0.06 | 12 | -0.76 |

The distribution of charges based on a NBO calculation at B3LYP/6-31G(d) level reveals that the carbon atoms are strongly negative. The hexagonal ring with the carbon atom in the centre contains boron atoms of types B2 and B3 with alternating small negative and positive charges.

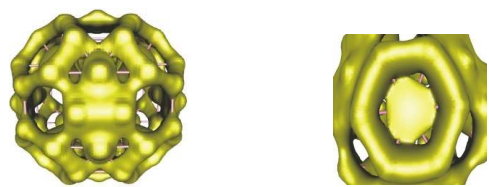

**a.** iso value 0.119    **b** iso value 0.107
**Fig. 4** Total density distribution

The boron cap atoms transfer nearly 2 electrons in the B4-B4 bond directions. Total density maps (Fig.4) show that the density of the methyne caps is nearly cylindrical with slight localisation in the direction of the B2 atoms.

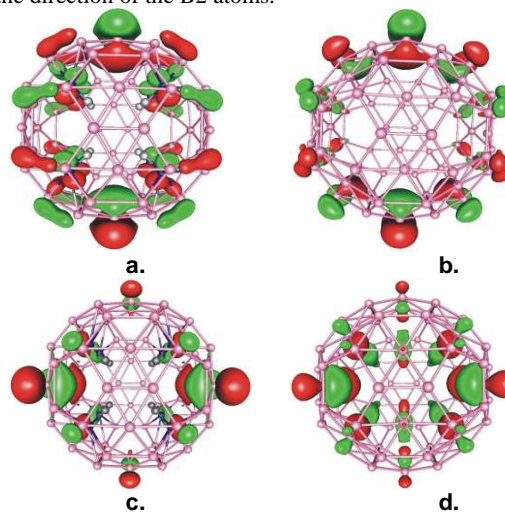

**Fig.5** Frontier orbitals of endo-72 and $B_{80}$: (a).HOMO of endo-72, (b.) HOMO of $B_{80}$, (c) LUMO of endo-72 , (d) LUMO +2 of $B_{80}$.

On the other hand there is a clear evidence for formation of a 4-centre bond between the B1 and B4 atoms. In total six such bonds can be realised in the structure with 12 boron caps. This will contribute to the stability of endo-72.

The HOMO has $t_u$ symmetry and has a shape similar to the HOMO of boron buckyball ( Fig. 5.a-b). The LUMO is also of $t_u$ symmetry but corresponds to the level LUMO +2 in $B_{80}$. (Fig 5.c-d). All these orbitals have π character like the boron buckyball and the original buckministerfullerene.[2-4]

**Table 4.** Symmetries and energies of valence MOs

| MOs | Energy (e.V) | MOs | Energy (e.V) | MOs | Energy (e.V) |
|---|---|---|---|---|---|
| 16 $t_g$ | -13.26 | 23 $t_u$ | -9.89 | 8 $a_u$ | -7.32 |
| 6 $a_u$ | -13.18 | 9 $e_g$ | -9.38 | 13 $a_g$ | -7.17 |
| 19 $t_u$ | -12.79 | 24 $t_u$ | -8.88 | 10 $e_g$ | -7.03 |
| 20 $t_u$ | -12.35 | 19 $t_g$ | -8.79 | 27 $t_u$ | -6.99 |
| 4 $e_u$ | -12.17 | 12 $a_g$ | -8.33 | 23 $t_g$ | -6.89 |
| 10 $a_g$ | -11.99 | 25 $t_u$ | -8.29 | 6 $e_u$ | -6.66 |
| 17 $t_g$ | -11.54 | 20 $t_g$ | -8.21 | 11$e_g$ | -6.58 |
| 21 $t_u$ | -11.51 | 7 $a_u$ | -7.94 | 28$t_u$ | -6.30 |
| 8 $e_g$ | -11.10 | 5 $e_u$ | -7.86 | *29$t_u$* | *-3.70* |
| 22 $t_u$ | -10.87 | 21 $t_g$ | -7.82 | *24$t_g$* | *-3.30* |
| 18 $t_g$ | -10.77 | 22 $t_g$ | -7.47 | *30$t_u$* | *-3.06* |
| 11 $a_g$ | -10.10 | 26 $t_u$ | -7.38 | | |

The 28 $t_u$ is the HOMO and the three last orbitals in italic are virtuals. (Table 4).

In conclusion methyne substitution of boron caps in the boron buckyball leads to a " viable" carbo-boron structure in the Hoffmann - Schleyer - Schaefer sense,[19] provided the remaining boron caps can be stabilized by 4-centre bonds as in the case for $B_{72}(CH)_8$. There is a strong endohedral pyramidalisation of the carbon atoms. The most interesting aspect about this new structure is that the hydrogen sites provide possibilities for further chemical modifications which may constitute possible synthetic routes.

We gratefully thank the Flemish Science Fund (FWO-Vlaanderen) and the K.U.Leuven Research Council for its continued financial support.

## Notes and references